\documentclass[%
reprint,
superscriptaddress,
footinbib,
amsmath,amssymb, 
aps,
prl
]{revtex4-2}
\usepackage{lipsum} 
\usepackage{graphicx}
\usepackage{dcolumn}
\usepackage{bm}
\usepackage[utf8]{inputenc}

\usepackage{lipsum} 
\usepackage{xcolor}

\newcommand{\ins}{\text{i}}
\newcommand{\out}{\text{o}}

\newcommand{\z}{c_0}

\newcommand{\PSIinf}{\psi^\infty}
\newcommand{\Rmean}{\langle R \rangle}
\newcommand{\psiaver}{\Psi_\text{tot}}
\newcommand{\sour}{s}

\newcommand{\io}{\text{i/o}}

\newcommand{\eq}[1]{Eq.~(\ref{eq:#1})}

\newcommand{\nn}{\nonumber}

\begin{document}
\author{Jonathan Bauermann}
\email{jbauermann@fas.harvard.edu}
\affiliation{Department of Physics, Harvard University, Cambridge, MA 02138, USA}
\author{Giacomo Bartolucci}
\affiliation{Department of Physics, Universitat de Barcelona, Barcelona, Spain}
\author{Christoph A. Weber}
\affiliation{Faculty of Mathematics, Natural Sciences, and Materials Engineering, Institute of Physics,
University of Augsburg, Augsburg, Germany}
\author{Frank Jülicher}%
\email{julicher@pks.mpg.de}
\affiliation{Max Planck Institute for the Physics of Complex Systems, Dresden, Germany}
\affiliation{Center for Systems Biology Dresden, Dresden, Germany}
\affiliation{
Cluster of Excellence Physics of Life, TU Dresden, 01062 Dresden, Germany 
}

\title{
Theory of Reversed Ripening in Active Phase Separating Systems
}
\date{\today}

\begin{abstract}
The ripening dynamics in passive systems is governed by the theory of Lifshitz-Slyozov-Wagner (LSW). 
Here, we present an analog theory for reversed ripening in active systems.
To derive the dynamic theory for the droplet size distribution, we consider a minimal ternary emulsion with one active reaction, leading to one conserved quantity.
Even for cases where single droplets constantly grow, coupling many droplets via the conserved density in the far field leads to a self-organized reversal of ripening and, thus, a monodisperse emulsion.
For late times, we find a scaling ansatz leading to the collapse of the rescaled size distributions, different from the LSW theory.
This scaling behavior arises from a stable fixed point in the single droplet dynamics and may capture the late-time behavior of many active matter systems exhibiting reversed ripening.
\end{abstract}

\maketitle

Domain ripening is a hallmark of the dynamics of phase-separated systems such as emulsions that undergo liquid-liquid phase separation or liquids undergoing crystallization~\cite{ostwaldStudienUberBildung1897,voorheesTheoryOstwaldRipening1985, pollock1994directional, bibette1999emulsions, ratkeGrowthCoarseningOstwald2002}. 
In such processes, larger domains of a new phase grow at the expense of smaller ones that shrink and eventually dissolve, allowing the system to approach thermodynamic equilibrium. 
The seminal work of Lifshitz, Slyozov, and Wagner revealed that in three-dimensional systems, the average droplet size scales with $\langle R \rangle(t) \propto t^{1/3}$ and the number of droplets decreases as $N(t)\propto t^{-1}$~\cite{wagnerTheorieAlterungNiederschlagen1961, lifshitzKineticsPrecipitationSupersaturated1961}. Furthermore, in the late stages of ripening, they reported a universal shape of the droplet size distribution that is independent of the physical parameters and time.

In active systems, the droplet dynamics can differ fundamentally. Ultimately, ripening can even be reversed when the system is actively maintained away from equilibrium~\cite{weberPhysicsActiveEmulsions2019, Brauns2021, Cates2024}.
Reversed ripening was found in 
simulations of active Brownian particles~\cite{stenhammar2014phase, patch2018curvature, yan2025stochastic} and in scalar field theories such as the active model  B+~\cite{tjhung2018cluster, fausti2024statistical} and extensions with hydrodynamic flows~\cite{Singh2019}.
Similarly, reversed ripening has been reported in chemically active emulsions~\cite{glotzerMonteCarloSimulations1994, zwickerSuppressionOstwaldRipening2015,wurtzChemicalReactionControlledPhaseSeparated2018}, where chemical reactions are maintained away from equilibrium~\cite{lee2018novel, weberPhysicsActiveEmulsions2019, zwicker2024chemicallyactivedroplets, julicher2024droplet}.

Combining phase separation and non-equilibrium chemical reactions, provides a paradigm relevant to cell biology, attested by a growing number of studies of biomolecular condensates involved in the spatial organization of biochemical processes in living cells~\cite{hymanLiquidLiquidPhaseSeparation2014, shinLiquidPhaseCondensation2017, bananiBiomolecularCondensatesOrganizers2017, boeynaemsProteinPhaseSeparation2018, nakashimaBiomolecularChemistryLiquid2019}.
Such chemically active emulsions have also gained a lot of interest in recent years, with experimental studies of
biophysical~\cite{strulsonRNACatalysisCompartmentalization2012, drobotCompartmentalisedRNACatalysis2018} and chemical systems~\cite{donauActiveCoacervateDroplets2020, 
schwarz2021parasitic, heckel2021spinodal,
donauPhaseTransitionsChemically2022}, as well as theoretical studies~\cite{wurtzChemicalReactionControlledPhaseSeparated2018, kirschbaumControllingBiomolecularCondensates2021, bauermannEnergyMatterSupply2022, 
kumarFluctuationsShapeDependence2023, choNonequilibriumInterfacialProperties2023, 
demarchi2023enzyme,
häfner2024reaction, zwicker2024chemicallyactivedroplets}.
Reversed ripening in binary, chemically active emulsions is driven by a simple mechanism~\cite{glotzerMonteCarloSimulations1994, zwickerSuppressionOstwaldRipening2015}: When the components enriched in the droplet are actively degraded inside or produced outside the droplet, a stationary droplet radius emerges where diffusive and chemical fluxes balance, defining a unique length scale. In multi-component mixtures, however, no such unique length scale appears; droplet size instead depends on kinetic rates of production, degradation, and conserved quantities~\cite{bauermann2024ripe}. How reversed ripening arises in these more general multi-component systems remains unclear.

Here, we consider a chemically active emulsion and show how the coupling among many droplets via conserved quantities collectively leads to a stable fixed point in the droplet dynamics.
Our theory builds on methods developed by Lifshitz, Slyozov, Wagner, and others, for binary emulsions without chemical reactions~\cite{wagnerTheorieAlterungNiederschlagen1961, lifshitzKineticsPrecipitationSupersaturated1961, voorheesOstwaldRipeningTwoPhase1992, yao1993}. 
In passive systems, the collective dynamics of coupling individual droplets in the far field is key to understanding ripening. Here, we show that in chemically active emulsions, this coupling can lead to a reversal of ripening and a new scaling for late times.

\begin{figure}[tb]
    \centering
    \includegraphics[width=0.47 \textwidth]{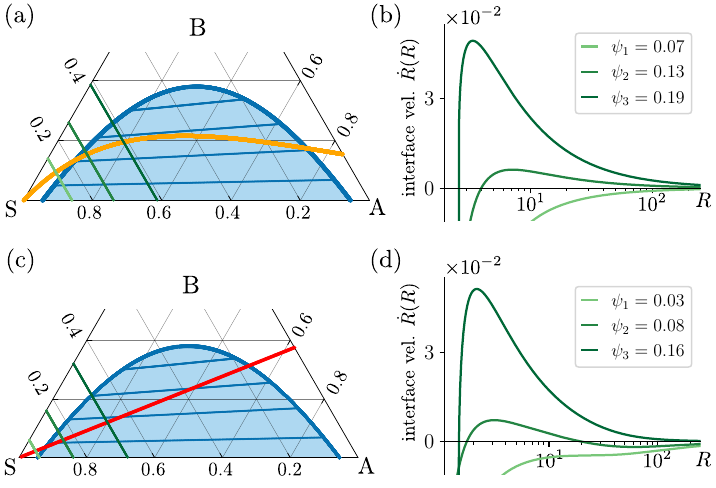}
    \caption{\label{fig:sketch} 
    \textbf{Passive and active emulsion.}
    (a,c): phase diagram of ternary emulsions with passive or active chemical reaction A $\rightleftharpoons$ B.
    The binodal (thick blue), tie-lines (thin blue), and the reaction nullcline for passive (yellow) and active (red) reactions are shown.
    Lines of constant $\psi = 0.07, 0.13, 0.19$ (a) and $\psi = 0.03, 0.08, 0.16$ (b) are highlighted (green).  
    (b,d): Interface velocities $\dot{R}$ as a function of radius $R$ for single droplets are shown for these values of $\psi$. 
    Phase diagram and interface velocities were obtained using a Flory-Huggins mean field model of mixtures; see the End Matter for details.}
\end{figure}

\textit{Theory of reversed ripening - }To develop such a general theory, we consider a chemically active emulsion as a model for phase-separated active matter undergoing reversed ripening. This framework incorporates both conserved densities and active chemical reactions. 
A minimal case is a ternary mixture with a single reaction: considering components $A$, $B$, and $S$ and the monomolecular reaction $A \rightleftharpoons B$. The concentration dynamics for $i = A, B$ are given by $\partial_t c_i = \Gamma_i \nabla^2 \mu_i  - k_{ji} c_i + k_{ij} c_j$, where $\Gamma_i$  are mobility coefficients, $\mu_i$ the chemical potentials determined by the system’s free energy, and $k_{ij}$ specify the reaction kinetics. 
For passive systems, the rates satisfy $k_{AB} c_B= k_{BA} c_A$ when $\mu_A = \mu_B$, leading to a unique thermodynamic equilibrium, as shown in the ternary phase diagram (Fig.~\ref{fig:sketch}(a)). Here, the reaction nullcline (yellow) intersects the binodal at points joined by a tie-line, allowing for a thermodynamic equilibrium. 
If $k_{ij}$ do not obey this relation, the system is chemically active — for example, if the rates are constant as in Fig.~\ref{fig:sketch}(c), the reaction nullcline (red) intersects the binodal at two unconnected points, preventing relaxation to equilibrium~\cite{bauermannChemicalKineticsMass2022}.

The coexistence of phases is governed by the quantity $\psi = (c_A +c_B)/2$ that is conserved by the chemical reaction,
both in the passive and the active 
case. We illustrate this property in Fig.~\ref{fig:sketch} with three different lines of constant $\psi$ shown in green in the phase diagrams for each case. 
The line of the smallest value of the conserved quantity  $\psi$ (brightest green) intersects the reaction nullcline outside the binodal domains, leading to a homogeneous steady state with vanishing reaction fluxes. However, for larger values of $\psi$ (darker green shades), this intersection lies within the binodal domain.
Thus, two phases coexist. 
For the passive case, there is a unique thermodynamic equilibrium state for which the coexisting phases and the chemical reactions are at equilibrium. 
A passive emulsion of droplets will undergo ripening, relaxing toward this thermodynamic equilibrium state.
However, the behavior on long-time scales of an active emulsion is unclear since thermodynamic equilibrium cannot act as a global attractor of the ripening dynamics.
To derive the dynamics of such an active emulsion, we express the dynamics of the conserved quantity $\psi = (c_A + c_B)/2$, together with the dynamics of the non-conserved quantity $\xi = (c_A - c_B)/2$, the extent of reaction, as
\begin{equation}
\partial_t \psi = \Gamma\nabla^2 \mu_\psi   \;, \; \;\;
\partial_t \xi = \Gamma \nabla^2 \mu_\xi -K \xi + \mathcal{S} \psi  \;,
 \label{eq:dyn_psi_xi} 
\end{equation}
where $K= k_{AB} + k_{BA}$, $\mathcal{S} = k_{AB} - k_{BA}$. We chose the same mobility $\Gamma=\Gamma_i$ for both components, for simplicity, and defined $\mu_\psi = (\mu_A + \mu_B)/2$ and $\mu_\xi = (\mu_A - \mu_B)/2$ as  the corresponding chemical potentials of $\psi$ and $\xi$.

\textit{Single droplet dynamics - }
We derive the slow interface dynamics in a quasistationary limit. To this aim, we determine the stationary solutions of the dynamical Eqs.~\eqref{eq:dyn_psi_xi} for one single droplet of radius $R$ located at the origin of the coordinate system. Specifically, we expand \eq{dyn_psi_xi} up to linear order on both sides, inside ($r< R$) and outside ($r>R$), around the respective values of concentrations at the interface, leading to
\begin{equation}
    \partial_t \psi = D_\psi^\io \nabla^2 \psi \;, \; \;\; \partial_t \xi = D_\xi^\io \nabla^2 \xi - k^\io \xi + s^\io  \psi \; ,
    \label{eq:rd_eq}
\end{equation} 
where $D_\psi$, $D_\xi$, $k$ and $s$ denote the kinetic coefficients and the upper indices $\io$ denote the inside/outside domain, respectively~\cite{bauermannEnergyMatterSupply2022}.
The stationary field for the conserved quantity is given by
\begin{gather}
    \psi^{\ins} = \Psi_{R}^\ins \; , \;\;\; \psi^{\out}(r) = \Psi^\infty + \frac{R\left(\Psi_R^\out -\Psi^\infty  \right)}{r} \; ,
\end{gather}
where $\Psi_{R}^\io$ denote the concentrations of $\psi$ at the interface and $\Psi^\infty$ the value far from the droplet. For the non-conserved field, we find
\begin{subequations}
\begin{align}
    \xi^{\ins}(r) &=\frac{s^\ins  \Psi_{R}^\ins}{k^\ins} + \left(\Xi_R^\ins - \frac{s^\ins  \Psi_{R}^\ins}{k^\ins} \right) \frac{i_0(\lambda^\ins r)}{i_0(\lambda^\ins R)}, \\
     \xi^{\out}(r) &= 
    \frac{s^\out \Psi^\infty}{k^\out}  +
    \frac{s^\out R\left(\Psi_R^\out -\Psi^\infty  \right)}{k^\out r}  \nn \\
    &+\left(\Xi_R^\out -\frac{s^\out\Psi_R^\out}{k^\out}  \right) \frac{k_0(\lambda^\out r)}{k_0(\lambda^\out R)}\label{eq:xiout_spat} \;.
\end{align}
\end{subequations}
Similarly, $\Xi_R^\io$ denote the concentrations of $\xi$ at both sides of the interface, $i_0(x)= \sinh(x)/x$, and $k_0(x)=  \exp(-x)/x$ are the modified spherical Bessel functions of the first and second kind and zeroth order. 
Furthermore, we have introduced the inverse reaction-diffusion length scale $\lambda^\io = \sqrt{k^\io/D_\xi^\io}$.

\begin{figure*}[!t]
    \centering
    \includegraphics[width=0.95 \textwidth]{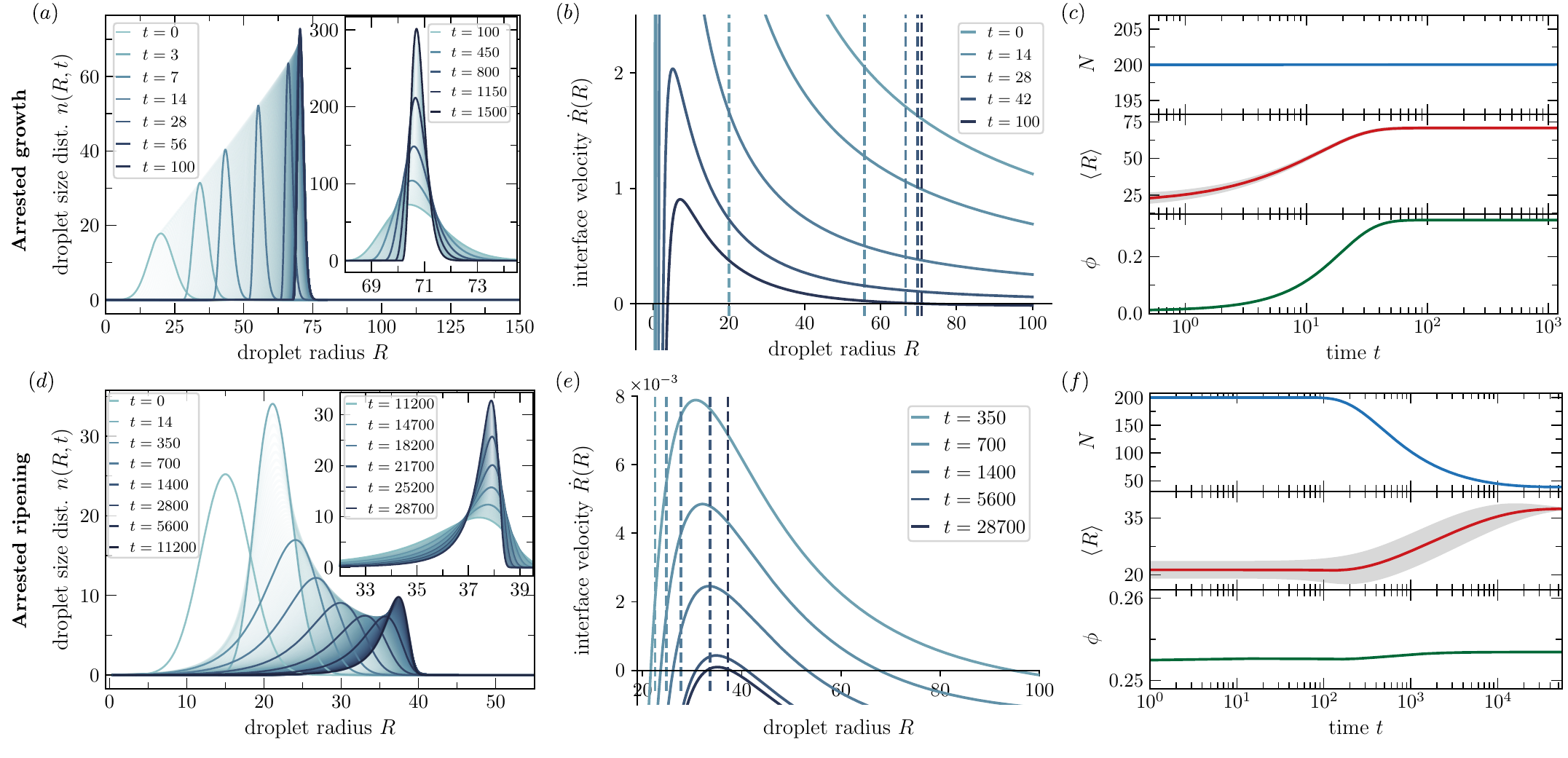}
    \caption{\label{fig:2example} 
    \textbf{Dynamics of droplet size distributions in chemically active emulsions.}
    (a) Distribution $n(R)$ of droplet sizes for different times $t$ for an arrested growth scenario (thick blue lines). The shading indicates distributions at intermediate times, revealing the arrest in a monodisperse distribution at late times. The inset shows late-time distributions.
    (b) Interface velocities $\dot{R}(R)$ for selected time points. The average radii at different times are indicated (dashed vertical lines)
    (c) Droplet number $N$, average radius $\langle R \rangle$, and the phase fraction $\phi$ as a function of time.
    (d-f) Same plots as (a-c) but for a scenario of arrested ripening.
    Parameters: Initial droplet number $N=200$ in reference volume $V_\text{ref}$, Gaussian initial distribution $\mathcal{N}(\bar{R}_0, \sigma_R) $,  $b=1$,  $c_0=1$, $D_{\psi/\xi}^\io = 1$, $\gamma = 1/6$, $V_\text{ref} = 4 \pi R_\text{ref}^3/3 $, $R_\text{ref}=600$, $k_{AB}^\io/k_{BA}^\io=2.5$; arrested growth: $\alpha = 0.15 \pi$, $\Psi_\text{tot}=-0.5$, $k_{AB}^\io=0.0013$, $\bar{R}_0= 20$, $\sigma_R = 20$, arrested ripening: $\alpha = 0.23\pi$, $\Psi_\text{tot}=-0.7$, $k_{AB}^\io=0.0009$, $\bar{R}_0 = 15$, $\sigma_R = 10$.
    }
\end{figure*}
To determine the remaining interface conditions, we impose that phase equilibrium is satisfied locally, assuming a slow-moving interface. Thus, the four concentrations $\Psi^\text{i/o}_R,\Xi^\text{i/o}_R$ at the interface of a droplet with radius $R$ have to satisfy
~\cite{safranStatisticalThermodynamicsSurfaces2019}
\begin{subequations}
 \label{eq:phase_eq} 
\begin{gather}
\mu_\psi^\ins = \mu_\psi^\out \;, \;\;\;\mu_\xi^\ins = \mu_\xi^\out \;, \\
f^\ins - f^\out = \mu_\psi^\io \left(\Psi_R^\ins -\Psi_R^\out  \right) + \mu_\psi^\io \left(\Xi_R^\ins -\Xi_R^\out \right) - \frac{2 \gamma}{R}\;,
\end{gather}
\end{subequations}
where $f^\io$ is the free energy density evaluated in the corresponding phase, $\mu_n^\io = \partial f^\io/ \partial n $, with $n=\psi,\xi$, are the corresponding chemical potentials, and $\gamma$ the surface tension of the interface.
Similar to ripening in ternary passive mixtures \cite{kuehmannOstwaldRipeningTernary1996}, the final interface condition follows from local material conservation:
\begin{gather}
    \frac{j_\psi^\ins(R) - j_\psi^\out(R)}{\Psi_R^\ins-\Psi_R^\out} = \frac{j_\xi^\ins(R) - j_\xi^\out(R)}{\Xi_R^\ins-\Xi_R^\out} \label{eq:delj} \;,
\end{gather}
where the flux densities in the radial direction defined are $j_\psi^\io (R) = - D_\psi^\io \partial_r \psi^\io(R)$ and $j_\xi^\io(R) = - D_\xi^\io \partial_r \xi^\io(R)$.
With the four conditions Eqs.(\ref{eq:phase_eq},\ref{eq:delj}), all four interface concentrations are uniquely defined.
Furthermore, the same local material conservation law at the interface sets its velocity $\dot{R} = (j_\psi^\ins(R) - j_\psi^\out(R))/(\Psi_R^\ins-\Psi_R^\out)$ ~\cite{brayTheoryPhaseorderingKinetics1994}. Thus, we find
\begin{equation}
    \dot{R}(R, \Psi^\infty)= \frac{D_\psi^\out}{R} \frac{\Psi^\infty-\Psi_R^\out (R, \Psi^\infty)}{\Psi_R^\ins(R, \Psi^\infty)-\Psi_R^\out(R, \Psi^\infty)} \;.\label{eq:dRdt}
\end{equation}
We have explicitly stated that the equilibrium values $\Psi_R^\io$  depend on both the droplet radius $R$ and the far-field concentration  $\Psi^\infty$, in contrast to the simple case of binary mixtures, where they only depend on $R$.

In Fig.~\ref{fig:sketch}(b,d), we show $\dot{R}$ as a function of radius $R$ for a single droplet for passive and active cases.
For conserved quantities that intersect the reaction nullcline within the binodal region (cases of $\psi_2$ and $\psi_3$ in Fig.~\ref{fig:sketch}), there exists one unstable root of the interface velocity, known as the critical nucleation radius. 
For passive systems, the interface velocity stays positive for larger droplets, and a single droplet larger than this radius continuously grows.
For active systems, however, there is a range of conserved quantities where a second root of the $\dot{R}$-curve exits, which is stable. This case corresponds to intensive droplets; see, for example, $\psi_2$ in Fig.~\ref{fig:sketch}(d). 
However, the $\psi$-range with intensive droplets is finite. At a critical value of the conserved quantity $\psi=\psi_\text{crit}$, the stable root diverges.
Beyond this critical transition, single droplets grow limitless, and are thus referred to as extensive droplets~\cite{bauermann2024ripe}. 

\textit{Emulsion dynamics - }
Following the approach developed for Ostwald ripening~\cite{wagnerTheorieAlterungNiederschlagen1961, lifshitzKineticsPrecipitationSupersaturated1961, voorheesOstwaldRipeningTwoPhase1992, yao1993}, we introduce the droplet size distribution $n(R,t)$. 
After droplets have nucleated, the initial distribution evolves in time according to a conserved dynamics:
\begin{equation}
    \partial_t n = - \partial_R (n \dot{R}) \;. \label{eq:dndt}
\end{equation}
The disappearance of droplets at $R=0$ is reflected by the boundary condition $n(R=0,t)=0$ for all times $t$. With the help of this distribution $n(R,t)$, we define the phase fraction 
\begin{equation}
    \phi(t) = \frac{4 \pi }{3 V_\text{ref}} \int_0^\infty dR \; R^3 n(R,t) \;, \label{eq:phi}
\end{equation}
which denotes the fraction of volume occupied by $ N(t) = \int_0^\infty dR \; n(R, t) $  droplets in the reference volume $V_\text{ref}$ at time $t$. The average droplet radius is
\begin{equation}
\langle R (t)\rangle = \frac{1}{N(t)} \int_0^\infty dR \, R \, n(R,t) \label{eq:averR} \;.
\end{equation}

Finally, we have to invoke the overall conservation of the conserved material $\psi$ in our system.  We write as a mean-field approximation
\begin{equation}
\Psi_\text{tot} = \phi (t) \,  \Psi_R^\ins \big(\langle R(t) \rangle, \Psi^\infty(t) \big) + \big(1-\phi(t)\big)  \, \Psi^\infty (t)\label{eq:psi_inf} \; , 
\end{equation}
where $\Psi_\text{tot}$ is the overall conserved concentration of $\psi$ in our reference system, which is fixed and serves as a control parameter. 

These equations provide a closed dynamics for the droplet size distribution. 
In the following examples, we choose for simplicity a  Ginzburg-Landau free energy of a ternary mixture, i.e., $f(c_A, c_B) = b(c^\text{c}_A +  c^\text{s}_B+ c_0)^2(c^\text{c}_A +  c^\text{s}_B- c_0)^2/(8 c_0^2) + b (c^\text{c}_B -  c^\text{s}_A)^2/2$ where $b$ is an energy scale, $c_0$ a reference concentration, and $c^\text{c}_i = \cos(\alpha) c_i$ and $c^\text{s}_i = \sin(\alpha) c_i$, thereby parameterizing the interactions between components by a compositional angle $\alpha$ \cite{bauermann2024ripe}; see the End Matter for further details.

\begin{figure*}[!t]
    \centering
    \includegraphics[width=0.95 \textwidth]{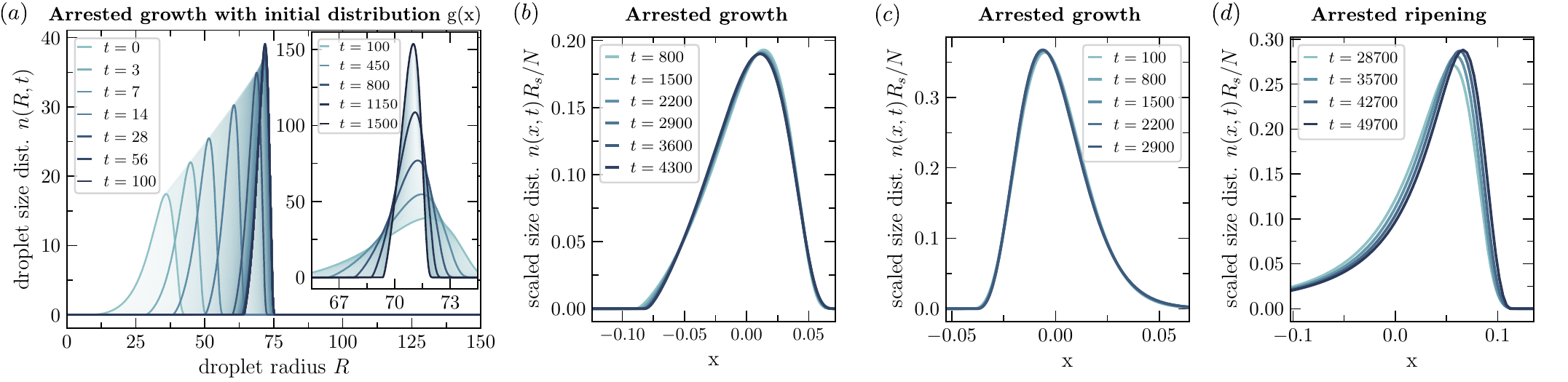}
    \caption{ \label{fig:scaling} 
    \textbf{Shape of droplet size distributions at late times.}
    (a) Dynamics of the droplet size distribution $n(R)$ for the same system as shown in Fig.~\ref{fig:2example}(a) but with an initial shape $g(x)$. 
    The inset shows the distribution at late times.
    (b) Late time distribution as shown in (a)(inset), but scaled and shown as a function of scaled radius $x$.
    (c,d) Scaled plot as in (b) but showing late time distributions of arrested growth (Fig.~\ref{fig:2example}(a)) and arrested ripening (Fig.~\ref{fig:2example}(d)).
    }
\end{figure*}

In Figs.~\ref{fig:2example}(a,d), we show two cases of the time evolution of the droplet size distribution as a function of the droplet radius $R$ for different values of $\alpha$, late times are shown in the insets.
For both examples, $\Psi_\text{tot}$ is chosen within the regime of extensive droplets, where single droplets grow without
bounds \cite{bauermann2024ripe}, and the initial density distribution is Gaussian with $N_0$ droplets in a reference volume $V_\text{ref}$. In the first case, all droplets grow initially and reach a monodisperse steady state; see Fig.~\ref{fig:2example}(c). We term this case arrested growth.
In the second case, droplets exhibit ripening at early times, with smaller droplets shrinking while larger ones grow, thereby reducing the droplet number $N$ (see Fig.~\ref{fig:2example}(f)). At later times, the ripening stops, and a monodisperse stationary state of $N<N_0$ droplets is reached. We term this case arrested ripening.
In both cases, the average droplet radius $\langle R \rangle$ grows initially and relaxes towards a constant value at large times; see Figs.~\ref{fig:2example}(c),(f).
This relaxation to a monodisperse state can be explained as follows: 
As droplets grow by the influx of conserved material, the dilute phase concentration $\Psi^\infty$ decreases.  As a result, a stable droplet radius emerges for $\Psi^\infty < \psi_\text{crit}$.
Once the average droplet radius $\langle R \rangle$ (indicated by dashed lines in Figs.~\ref{fig:2example}(b,e)), which increases in time, meets the stable droplet radius, growth stops, and all droplets converge to the stable droplet radius. The result is a monodisperse emulsion.
For comparison, in the case of Ostwald ripening of passive mixtures, the average droplet radius is always equal to the unstable droplet radius while no stable droplet radius exists, and the size distribution broadens.


\textit{Steady state - }
We can characterize the properties of the steady state of the active emulsion. If the number of droplets $N$ is given, the size distribution is $n(R,\infty) = N \delta (R-R_\text{s})$, where $R_\text{s}$ denotes the stationary radius which obeys
\begin{equation}
    \Psi^\infty_\text{s} = \Psi_R^\out(R_\text{s}, \Psi^\infty_\text{s}) \;.\label{eq:inter_stationarity}
\end{equation}
This equation, together with the lever rule for the conserved quantity
\begin{equation}
    \frac{4 \pi R_\text{s}^3 N_\text{s} }{3 V_\text{ref}} = \frac{\Psi_\text{tot} - \Psi^\infty_\text{s}}{\Psi_R^\ins(R_\text{s}, \Psi^\infty_\text{s})-\Psi^\infty_\text{s}} \;,
    \label{eq:infinity_phi}
\end{equation}
determine the unknown values of $R_\text{s}$ and $\Psi^\infty_\text{s}$.
Note that this lever rule depends on the stationary droplet density $N_\text{s}/V_\text{ref}$, in contrast to a passive system. Thus, many different solutions with different numbers of droplets exist.
Consequently, both the dynamics and the stationary state depend on the diffusivities and reaction rates, see End Matter and the supplementing movie for further details~\footnote{Note1}.

\textit{Scaling regime for late times - }
In passive mixtures, independently of the initial distribution, the size distribution widens and exhibits self-similarity at late times. There, $n(R,t) \propto  g(x) (\tau/t) \langle R(t) \rangle^{-1}$, where $\tau$ is a time-scale and $g(x)$ is a universal distribution that depends on the dimensionless variable $x= R/\langle R(t) \rangle$ \footnote{$g(x)= x^2 (3+x)^{-7/3}(3/2-x)^{-11/3}\exp(-x/(3/2-x))$}. The average radius scales $\langle R(t) \rangle \propto t^{1/3}$, and consequently $N(t) \propto t^{-1}$~\cite{wagnerTheorieAlterungNiederschlagen1961, lifshitzKineticsPrecipitationSupersaturated1961}.
In contrast, we find that in an active emulsion, the size distribution $n(R,t)$ at late times becomes narrow and centered around $R_\text{s}$. 
Droplets relax towards the stable radius $R_\text{s}$ according to $\dot{R} = - \lambda (R-R_\text{s})$, where $\lambda = - d \dot{R}/d R|_{R=R_\text{s}}$ using Eq.~\eqref{eq:dRdt}.
This implies that the size distribution evolves as
\begin{equation}
 \label{eq:ansatz} 
n(R,t) =  \frac{f(x)}{R_\text{s}} \text{e}^{\lambda t} \;, \;\;\;\;
x = \left(\frac{R - R_\text{s}}{R_s}\right) \text{e}^{\lambda t} \;,
\end{equation} 
where $f(x)$ is a time-independent function describing the shape.
Indeed, Eq.~\eqref{eq:ansatz} satisfies Eq.~\eqref{eq:dndt} in the long time limit for any normalizable function $f(x)$. Thus, $f(x)$ is not further constrained and depends on the initial distribution. 
Consequently, the droplet number is constant in time $N = \int_0^\infty dR \, n(R,t) = \int_0^\infty dx \, f(x) $.

To illustrate that the shape $f(x)$ of the size distribution depends on the initial distribution, we initialize the system presented in Figs.~\ref{fig:2example}(a-c) with the same number of droplets $N_0$ but with sizes following the universal distribution $g(x)$ as an initial condition.
Fig.~\ref{fig:scaling}(a) shows the droplet growth for this initial condition as well as growth arrest at late times (inset). The rescaled curves $n R_\text{s}/N$ as a function of scaled distance $x$ for late times are shown in Fig.~\ref{fig:scaling}(b).
For arrested growth, the total number of droplets is almost conserved, leading to the same. Thus, the stationary droplet size $R_\text{s}$ and the relaxation rate $\lambda$ are the same as for the previous case of arrested growth.
To reveal this difference in the rescaled distributions, we show the scaled size distribution at different late times for arrested growth shown in Fig.~\ref{fig:2example}(a) in Fig.~\ref{fig:scaling}(c). 
These scaled distributions converge to different shape functions $f(x)$ for late times; compare Fig.~\ref{fig:scaling}(b) and Fig.~\ref{fig:scaling}(c).
Furthermore, we show the scaled distribution for arrested ripening, as shown in Fig.~\ref{fig:2example}(d), in Fig.~\ref{fig:scaling}(d). Note that here, the shape function $f(x)$ of arrested ripening is similar to $g(x)$.

Finally, we comment on the effect of fluctuations on the droplet size distribution. In the simplest form, fluctuations can be accounted for by adding an effective diffusion term with diffusivity $\mathcal{D}$ to Eq.~\eqref{eq:dndt}, writing $
    \partial_t n = \mathcal{D} \partial^2_{R} n - \partial_R (n \dot{R})$.
Expanding the interface velocity around the stationary radius, $\dot{R} = -\lambda (R-R_\text{s})$, we obtain a Gaussian stationary distribution $n(R) =N \sqrt{\lambda/(2 \pi \mathcal{D})} \exp[-\lambda (R-R_\text{s})^2/(2 \mathcal{D})]$. Recently, it was shown that the size distribution of condensates driven by active chemical processes in the nucleolus of living cells, i.e., the nucleolar fibrillar center, is well matched by a Gaussian in the regime of arrested ripening \cite{Banani2024}, consistently with our theory.

\textit{Discussion - }
We developed a theory for reversed ripening that can emerge in chemically active emulsions. This theory uses a similar framework as the classical ripening of passive emulsions by Lifshitz, Slyozov, and Wagner~\cite{wagnerTheorieAlterungNiederschlagen1961, lifshitzKineticsPrecipitationSupersaturated1961}, however, the resulting dynamics for active emulsions is fundamentally different. 
In passive systems, the droplet size distribution broadens and reaches a universal shape when properly rescaled. 
In active emulsions with reversed ripening, we show that the distribution narrows, its shape depends on initial conditions, but a universal scaling behavior exists that leads to the collapse of the rescaled size distributions at late times.

Our theory naturally generalizes to more components and reactions, where the principles discussed here also apply. 
Our work could be applied to understand collective droplet dynamics in various experimental systems, such as biological condensates~\cite{Saurabh2022, Banani2024} or engineered active emulsions in synthetic chemistry~\cite{Nakashima2021, Donau2022}.
Several questions remain open in such active droplet systems, including the role of fluctuations, effects of droplet division, shape instabilities, droplet spacing, and nucleation. 
Integrating these and other aspects into our theory remains a challenge for future work.

Finally, reversed ripening has been reported in various other active matter systems,
including active Brownian particles~\cite{stenhammar2014phase, patch2018curvature, yan2025stochastic} and scalar field theories~\cite{tjhung2018cluster, fausti2024statistical, Singh2019}.
The scaling Ansatz Eq.~\eqref{eq:ansatz} considered in our theory for chemically active emulsions reflects the existence of a stable fixed point in droplet size in the emulsion's late-time dynamics.
Since stable fixed points in droplet size also exist in many other active matter systems~\cite{Woolley2010,Cates2024}.
We expect the validity of the scaling Ansatz and that the droplet size distribution evolves according to Eq.~\eqref{eq:dndt} (without fluctuations).
More than 60 years ago, the theory of Lifshitz, Slyozov, and Wagner led to the discovery of the universal ripening dynamics of passive systems. The theory proposed in our work may provide a stepping stone to identify the universal features of reversed ripening in other active systems.

\textit{Acknowledgment - }
J.B. thanks A. Goychuk for an insightful discussion and the German Research Foundation for financial support through the DFG Project BA 8210/1-1. 
G.B. thanks the Agencia Estatal de Investigación for funding through the Juan de la Cierva postdoctoral programme JDC2023-051554-I.
C.\ Weber acknowledges the European Research Council (ERC) for financial support under the European Union’s Horizon 2020 research and innovation programme (``Fuelled Life'' with Grant agreement No.\ 949021).

\newpage
\clearpage
\onecolumngrid
\section{End Matter}
\subsection{Multicomponent Flory-Huggins model}
We used a simple Flory-Huggins free energy density of a ternary mixture for generating Fig.~1. This free energy is typically written in terms of the volume fractions $\phi_A$, $\phi_B$ and $\phi_S$, i.e., the fraction of space occupied by particles of type $A$, $B$ and $S$, such that $\phi_A+\phi_B +\phi_S =1$. In its simplest form, it reads
\begin{equation}
    f(\phi_A, \phi_B) = \frac{k_B T}{\nu} \left( 
    \sum_{i=A,B,S} \phi_i \log(\phi_i) + 
    \sum_{i,j=A,B,S}\chi_{ij}\phi_i \phi_j +
    \sum_{i=A,B,S}w_i\phi_i
    \right) ,
\end{equation}
where $k_B T$ sets the energy scale and $\nu$ is a reference volume of the molecules. The first sum accounts for the entropic contribution of mixing. The second sum represents an energy contribution coming from the interaction between components $i,j$. Here, $\chi_{ij}$ is a dimensionless interaction strength. Furthermore, it is sufficient to take only the pairs $(i,j)=(A,B), (A,S), (B,S)$ into account. The last sum represents the internal energy contribution of molecules of type $i$, where $w_i$ is a dimensionless weight factor.

In the figure, we have chosen $\chi_{AB} = 0$, $\chi_{AS} = 3.2$, $\chi_{BS} = 0.4$, $w_{A} = w_{S} = 0$, $w_{B} = 1.8$, while the energy scale $k_B T$ and the reference volume $\nu$ do not affect the shape of the phase diagram at thermodynamical equilibrium.
However, for finite-sized droplets and their $\dot{R}$-curves, we have to specify the surface tension $\gamma$: we have chosen $\gamma \beta \nu  = \ell/6 $, where $\ell$ is our reference length scale, related to the interface width and sometimes called capillary width. In addition, we give all time scales in terms of $\tau$, the time it takes for a particle to diffuse across an area of $\ell^2$. Consequently, we set $D = 1 \ell^2/ \tau$. For the passive system, the chemical reaction finds its equilibrium fast and only the conserved quantity has to be taken into account. Therefore, to obtain the interface velocities, we have to replace Eq.~(6) with the condition $\mu_A^\io=\mu_B^\io$. The reaction rates, however, do not matter. For the active setting, however, we have to specify the rates. We set $k_{AB} = 0.001 \tau^{-1}$ and $k_{BA} = 0.0006 \tau^{-1}$, such that $k_{BA}/k_{AB} = 1/6$, the slope of the reaction nullcline in the phase diagram.

\subsection{Generic form of the free energy with simple phase equilibria}
Fig.~1 in the main text shows the $\dot{R}$, which is the key ingredient of the theory obtained from a Flory-Huggins free energy model, demonstrating that the theory proposed in the main text can be applied, in principle, to many classical problems studied in the literature. For such systems, however, the interface conditions (i.e., Eq.~(5)) lead to non-algebraic constraints that must be found numerically.
To simplify the analytics for the remaining figures and results, we have chosen a simpler free energy density combining a bi-quadratic and a quadratic term, i.e.,
\begin{align}
\label{eq:f}
f(c_A, c_B;\alpha) &= \frac{b}{8 c_0^2} (\cos(\alpha) c_A + \sin(\alpha) c_B + c_0 )^2 (\cos(\alpha) c_A + \sin(\alpha) c_B - c_0 )^2+ \frac{b}{2} (\cos(\alpha) c_B - \sin(\alpha) c_A )^2 \;,
\end{align}
where $b$ is a scale of the free energy density. 
For $\alpha=0$, the first term leads to a Cahn-Hilliard dynamics for $A$ with the equilibrium concentrations $\pm c_0$, while $B$ diffuses freely in and between the phases. Whenever $\alpha\neq 0$, molecular interactions rotate this simple free energy. For further details, see \cite{bauermann2024ripe}.

We can find the four equilibrium concentrations $\Psi^\io$ and $\Xi^\io$ analytically. Three conditions follow from the assumption of local phase equilibrium, i.e., Eq.~(5), while Eq.~(6) determines then uniquely one phase equilibrium as a function of $R$ and $\psi^\infty$. 
The Laplace pressure in Eq.~(5) ($\propto 2\gamma /R$), leads to a finite size correction of the equilibrium concentrations. By using a so-called Gibbs-Thomson coefficient $\beta$, we can approximate to linear order
\begin{align}
    \Psi^\io = \Psi^\io_0 + 2 \beta \gamma (\cos(\alpha)+\sin(\alpha))/R,  \;\;\; \Xi^\io = \Xi^\io_0 + 2 \beta \gamma (\cos(\alpha)-\sin(\alpha))/R, 
\end{align}
where $\Psi^\io_0$ and $\Xi^\io_0$ are the equilibrium values of an infinitely large system. These boundary conditions, together with Eq.~(6), determine uniquely the phase equilibrium of a droplet of size $R$.
The phase equilibria of the conserved quantity on both sides at the interface read
\begin{align}
    \Psi^\ins(R,\PSIinf) &=  
    \Big(k \PSIinf + \lambda (4 \beta \gamma k - \z R (k + \sour)) \cos(\alpha) - 
    \PSIinf \sour \cos(2 \alpha) + \lambda \z R (k - \sour) \cos(3 \alpha) - 
   2 (k \lambda ( \z R-2 \beta \gamma ) \nonumber \\
   & +  k \PSIinf \cos(\alpha) + 
      \lambda \z R (k + \sour) \cos(2 \alpha)) \sin(\alpha) + 
   2 k \lambda (2 \beta \gamma - \z R) \coth(\lambda R) (\cos(\alpha) + \sin(\alpha))\Big)  \nonumber \\
   &/\Big(k + 
   k \lambda R + (\lambda R-1) \sour \cos(2 \alpha) + 
   k ( \lambda R-1) \sin(2 \alpha) + 
   \lambda R \coth(\lambda R) (k + \sour \cos(2 \alpha) + k \sin(2 \alpha))\Big)
    \\
    \Psi^\out(R,\psi^\infty) &=
    \Big(k \PSIinf + (4 \beta \gamma k \lambda + (k + 2 k \lambda R - \sour) \z) \cos(\alpha) - 
   \PSIinf s \cos(2 \alpha) + (k - \sour) \z \cos(3 \alpha) \nonumber \\
   & - 
   2 (-k \lambda (2 \beta \gamma + R \z) + 
      k \PSIinf \cos(\alpha) + (k + \sour) \z \cos(2 \alpha)) \sin(\alpha) + 
   2 \lambda \coth(
     \lambda R) (\cos(\alpha) + \sin(\alpha)) \nonumber\\
     &\times 
     (2 \beta \gamma k + 
      R \sour \z \cos(2 \alpha) + k R \z \sin(2 \alpha))\Big) /\Big(k + 
   k \lambda R + (\lambda R-1) \sour \cos(2 \alpha) + 
   k ( \lambda R-1) \sin(2 \alpha) \nonumber \\
   & + \lambda R \coth(\lambda R) (k +\sour \cos(2 \alpha) + k \sin(2 \alpha))\Big), 
\end{align}
where $\lambda = \sqrt{k/D}$. Here, for simplicity, we have assumed $D_\psi^\ins = D_\psi^\out  = D_\xi^\ins = D_\xi^\out= D$ ,  $k^\ins = k^\out =  k $, $\sour^\ins = \sour^\out =  \sour $. With these solution and Eq.~(11), we find 
\begin{align}
\PSIinf(\Rmean, \phi) &= 
\Big(-k \psiaver (1 + \lambda \Rmean) + 
     \lambda \phi (4 \beta \gamma k -\Rmean (k + \sour) \z) \cos(\alpha) + 
     \psiaver (1 - \lambda \Rmean) \sour \cos(2 \alpha)+ 
     \lambda \phi \Rmean (k - \sour) \z \cos(3 \alpha) \nonumber\\
     & + 
     \lambda \phi (4 \beta \gamma k + \Rmean (\sour-k)\z) \sin(\alpha) - 
     \lambda \coth(\lambda \Rmean) (\cos(\alpha) + 
        \sin(\alpha)) (2 k \phi (-2 \beta \gamma + \Rmean \z) \nonumber \\
        &+ 
        \psiaver \Rmean (k + \sour) \cos(\alpha) + 
        \psiaver \Rmean (k - \sour) \sin(\alpha)) + 
     k \psiaver (1 - \lambda \Rmean) \sin(2 \alpha) - 
     \lambda \phi \Rmean (k + \sour) \z \sin(3 \alpha)\Big) \nonumber \\
     & 
     /\Big((1 + \lambda ( \phi-1) \Rmean) \sour \cos(2 \alpha) + 
     \lambda ( \phi-1) \Rmean \coth(\lambda \Rmean) (k + \sour \cos(2 \alpha) + k \sin(2 \alpha)) \\
     & 
     + k (-1 + \lambda (\phi-1) \Rmean + (1 + \lambda ( \phi-1) \Rmean) \sin(2 \alpha))\Big),
\end{align}
where $\psiaver$ is the concentration of the conserved density in the system.
With the equations above, we have explicit conditions for the boundary conditions and the closed set of dynamical equations, as described in the main test, can be iterated straightforwardly. 

\begin{figure*}[!t]
    \centering
    \includegraphics[width=0.65 \textwidth]{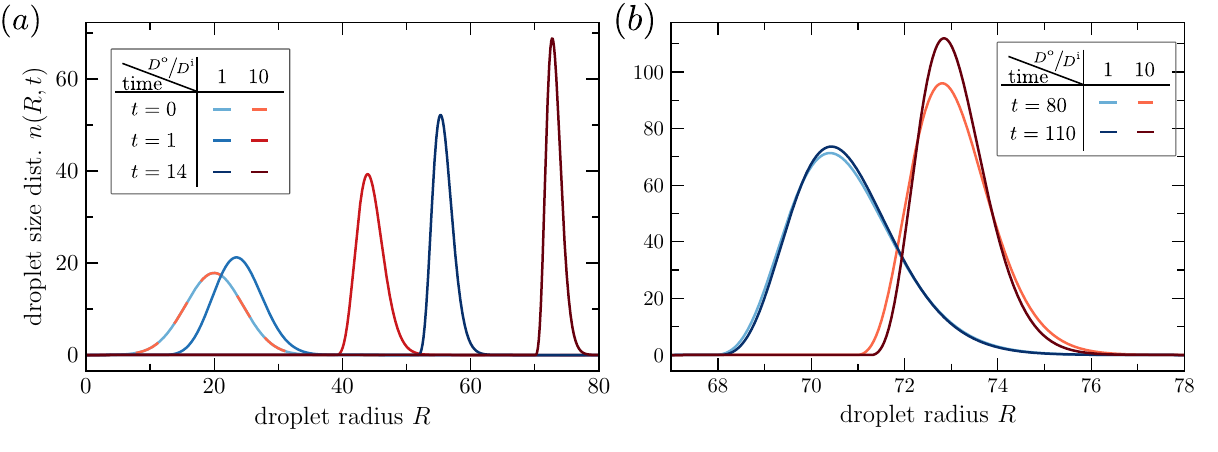}
    \caption{ \label{fig:SI_dif} 
    \textbf{Effect of diffusivities inside ($D^\text{o}$) and outside ($D^\text{i}$) of the droplet  on the droplet size distribution $n(R,t)$.}
    Early times (a) and late times (b) of the same dynamics shown in Fig.~\ref{fig:2example}(a-c) (blue) in comparison to the same system with a ten-times faster diffusion for both species $A$ and $B$ in the phase outside the droplet ($D^\text{o}=10 \, D^\text{i}$).
    For simplicity, different species have equal diffusivity in each phase. 
    For a movie of these dynamics, see the supplementary information.
    We see that the enhanced diffusivity outside speeds up the relaxation of the droplet size distribution to the state of arrested growth, while the final stationary distribution is less affected by this change of diffusivity. 
    }
\end{figure*}
The aforementioned equations can also be generalized for cases with different diffusivities and reaction rates in both phases. Here, we show the effect of faster diffusion in the dilute phase for the case of arrested growth shown in the main text.
We compare the dynamics of the system shown in Fig.~\ref{fig:2example}(a-c) (blue) to the same system, with a ten times faster diffusion in the outside phase in Fig.~\ref{fig:SI_dif}. In the early dynamics, see Fig.~\ref{fig:SI_dif}(a), faster diffusion allows the droplet size distribution to reach larger values of $R$ more quickly. Note, the stationary droplet sizes differ for the different diffusivities, as seen in the late dynamics in Fig.~\ref{fig:SI_dif}(b).

\end{document}